\documentclass[journal]{IEEEtran}

\usepackage{cite}
\usepackage[pdftex]{graphicx}
\usepackage{amssymb}
\usepackage[cmex10]{amsmath}
\usepackage{amsmath,amsfonts,amssymb}
\usepackage{algorithmic}
\usepackage{algorithm}
\usepackage{array}
\usepackage{textcomp}
\usepackage{lineno}
\usepackage{subfigure}
\usepackage{xcolor}
\usepackage{multirow}
\usepackage{diagbox}
\usepackage{threeparttable}
\usepackage{booktabs}
\usepackage{float}
\usepackage{stfloats}
\usepackage{url}
\usepackage[hidelinks]{hyperref}
\usepackage{verbatim}
\usepackage{graphicx}

\begin{document}
	
\title{AI Empowered Channel Semantic Acquisition\\for 6G Integrated Sensing and\\Communication Networks}

\author{
	Yifei Zhang, Zhen Gao, Jingjing Zhao, Ziming He, Yunsheng Zhang, Chen Lu, Pei Xiao
	
	\thanks{Y. Zhang and Z. Gao ({\em corresponding author}) are with
		the Advanced Research Institute of Multidisciplinary Science, Beijing Institute of Technology, Beijing 100081, China (email:
		\{zhangyifei22, gaozhen16\}@bit.edu.cn).}
	\thanks{J. Zhao is with
		the School of Electronics and Information Engineering, Beihang University, Beijing 100191, China, and also with the National Key Laboratory of CNS/ATM, Beijing 100191, China (email:
		jingjingzhao@buaa.edu.cn).}
	\thanks{Ziming He is with
		the Samsung Cambridge Solution Centre, System LSI, Samsung Electronics, Cambridge, U.K. (email:
		ziming.he@samsung.com).}
	\thanks{Yunsheng Zhang and Chen Lu are with
		the Shenzhen Institute of Information Technology, 518109, China (email:
		\{zhangys, luche\}@sziit.edu.cn).}
	\thanks{Pei Xiao is with
		the 5GIC \& 6GIC, Institute for Communication Systems (ICS), University of Surrey, GU2 7XH, UK (email:
		p.xiao@surrey.ac.uk).}
}


\maketitle

\begin{abstract}

Motivated by the need for increased spectral efficiency and the proliferation of intelligent applications, the sixth-generation (6G) mobile network is anticipated to integrate the dual-functions of communication and sensing (C\&S). Although the millimeter wave (mmWave) communication and mmWave radar share similar multiple-input multiple-output (MIMO) architecture for integration, the full potential of dual-function synergy remains to be exploited. In this paper, we commence by overviewing state-of-the-art schemes from the aspects of waveform design and signal processing. Nevertheless, these approaches face the dilemma of mutual compromise between C\&S performance. To this end, we reveal and exploit the synergy between C\&S. In the proposed framework, we introduce a two-stage frame structure and resort artificial intelligence (AI) to achieving the synergistic gain by designing a joint C\&S channel semantic extraction and reconstruction network (JCASCasterNet). With just a cost-effective and energy-efficient single sensing antenna, the proposed scheme achieves enhanced overall performance while requiring only limited pilot and feedback signaling overhead. In the end, we outline the challenges that lie ahead in the future development of integrated sensing and communication networks, along with promising directions for further research.

\end{abstract}

\IEEEpeerreviewmaketitle

\section{Introduction}\label{S1}

\IEEEPARstart{T}{he} future sixth-generation (6G) mobile communication network is expected to support a wide range of intelligent applications and services, necessitating not only increased network capacity, transmission rate, and reduced latency but also advanced sensing capability.
Among numerous studies, integrated sensing and communication (ISAC) stands out as one of the most promising and potential technologies to meet the aforementioned demands effectively \cite{6G-DFRC, JCR_in_PMN, 6G-JCS, 6G-ISAC}. Meanwhile, artificial intelligence (AI) presents opportunities for achieving synergy between communication and sensing (C\&S). The recently introduced channel semantic, in particular, offers a new perspective for ISAC signal processing \cite{CSAB}.

With the increasing demand for more spectrum resources, millimeter wave (mmWave) technology has been exploited in mobile communication. To compensate for the large path loss in mmWave wireless transmission, the beamforming gain provided by massive multiple-input multiple-output (MIMO) is essential. Nevertheless, the requirements of a large number of radio frequency (RF) chains in mmWave MIMO lead to prohibitive hardware costs and energy consumption. As a solution, the hybrid analog-digital (HAD)-MIMO has emerged, connecting a large number of antennas with only a few RF chains. With careful design of the phase shifter network or switches between RF chains and antennas, the HAD-MIMO can achieve performance comparable to the fully-digital MIMO, while significantly reducing hardware costs and energy consumption. Meanwhile, the concept of phased-MIMO radar, adopting the HAD-MIMO architecture, emerges to achieve a tradeoff between phased-array and MIMO radars \cite{6G-DFRC}. This similarity in the hardware design allows C\&S to be integrated into the same devices and systems. Besides, C\&S also exhibit many similarities in signal processing, such as beam training and target searching, beamforming and target tracking. These shared characteristics create additional possibilities for accomplishing C\&S using the same resources.

Over the last few decades, the coexistence of C\&S systems has been extensively investigated. These studies have primarily concentrated on interference control, with the aim of enabling two independently deployed systems to operate without interfering with each other. Although C\&S are co-located or even physically integrated, they still transmit different signals. These signals may partially overlap in time, frequency and/or spatial domains, sharing the same resources while attempting to minimize mutual interference as much as possible. However, effective interference cancellation imposes many stringent requirements, which severely limits its efficiency improvement in practice \cite{JCR_in_PMN}.

Since interference arises from the transmission of seprate signals, a natural approach to address this problem is to utilize a unified waveform for both C\&S. Nonetheless, this approach presents new challenges.
To begin with, the design of ISAC signal requires attention. Communication aims at improving the efficiency and reliability of data transmission, demanding high spectral efficiency and the capability to combat channel fading. While sensing requires good ambiguity function, large signal bandwidth, and long coherent processing interval to achieve high-resolution target sensing. Hence, the signal design of ISAC requires a balance between the performance of C\&S \cite{6G-JCS, 6G-ISAC}.
Following this, the ISAC signal processing emerges as a consideration. Although processing methods dedicated to sensing or communication might remain applicable, they fall short of effectively exploiting the benefit of mutual information. Additionally, the exploration of the correlation between C\&S channels is still in its early stages, lacking a unified paradigm to steer the development of integrated design. Furthermore, the novel characteristics of new ISAC signals pose challenges for achieving effectiveness and precision in the design.

The rest of this article is organized as follows. 
In the following sections, we first provide an overview of the state-of-the-art ISAC schemes and summarize their limitations. Then, we propose a novel deep learning (DL)-based channel semantic acquisition framework, which contains frame structure, pilot, and algorithm design to enable synergy between C\&S with extremely low costs.
Before concluding, important open research issues and future directions are discussed.

\section{Overview of the State-of-The-Art}\label{S2}

Waveform design and signal processing play key roles in ISAC systems. Specifically, waveform design focuses on achieving C\&S using shared resources. Depending on the type of the primary waveform, it can be categorized into sensing-centric, communication-centric, and joint design. Signal processing encompasses the encoding and decoding of communication symbols, as well as the detection and estimation of sensing targets. In this section, we present a comprehensive investigation of state-of-the-art waveform design and signal processing schemes, followed by a discussion of their limitations.

\subsection{Waveform Design}

While utilizing classical waveforms such as orthogonal frequency division multiplexing (OFDM) and chirp signals can accomplish C\&S tasks \cite{IEEE802, JRCD-SOTA}, a well designed waveform tailored for ISAC typically achieve superior performance \cite{SI-DFT-s-OFDM, CS-ISAC, WLAN-JCR, DFRC-OWD}.
In \cite{SI-DFT-s-OFDM}, a sensing-integrated discrete Fourier transform spread orthogonal frequency division multiplexing (SI-DFT-s-OFDM) system is proposed. This waveform provides lower peak-to-average power ratio than OFDM and integrates sensing through a frame structure with data and reference blocks. Besides, it is robust to delay spread with a flexible guard interval scheme.
Combining with compressive sensing (CS) techniques, \cite{CS-ISAC} designs a frame structure and a waveform tailored for the HAD-MIMO. The proposed scheme can adapt to fast time-varying environments and serve high-mobility user equipment (UEs).
Based on an ISAC of mmWave wireless local area network, \cite{WLAN-JCR} proposes an adaptive virtual waveform design. It achieves a tradeoff between C\&S performance, using the Cramér-Rao bound metric for sensing and a novel distortion minimum mean square error metric for data communication.
In \cite{DFRC-OWD}, globally optimal closed-form solutions for both omnidirectional and directional beampattern design problems are provided. Expanding on this groundwork, a low-complexity Riemannian conjugate gradient algorithm is proposed to flexibly balance C\&S performance. In order to address the challenges posed by practical constraints, \cite{DFRC-OWD} also introduces a globally optimal algorithm and provides an analysis of its worst-case complexity.

Despite the efforts mentioned above, there still exist several critical issues in waveform design:
\begin{itemize}
	\item The performance of waveform designed in Doppler and delay domains has received significant attention, while the angular domain, which is the most important domain in multi-antenna C\&S channels, appears to have been overlooked.
	
	\item Although efforts in waveform design have taken into account the balance of ISAC performance, design of the frame structure is relatively inflexible and fails to achieve synergy.
	
	\item There remains potential for enhancing the extraction of C\&S information, particularly concerning the utilization of prior information about specific channels to further optimize waveform designs.
\end{itemize}

\subsection{Signal Processing}

Signal processing for ISAC based on multi-carrier multi-user HAD-MIMO systems is challenging \cite{QoS, JRCD-SOTA, CS-ISAC}.
In \cite{QoS}, communication symbols are modulated in radar pulses with hybrid beamforming, and the quality of service requirements are considered. Furthermore, the multiple signal classification (MUSIC)-based angle of arrival (AoA) estimator is designed under the HAD-MIMO.
A framework capable of simultaneously target detection and channel estimation (CE) is introduced in \cite{JRCD-SOTA}. The proposed hybrid beamforming scheme can simultaneously communicate with UEs and track targets.
In order to improve the resolution of sensing while reducing hardware costs and energy consumption, a widely spaced array (WSA) with low-resolution analog-to-digital converters (ADCs) is considered in \cite{CS-ISAC}. The proposed CS-based scheme effectively overcomes angular ambiguity brought by the WSA.
Relying on the SI-DFT-s-OFDM waveform, a DL-based ISAC receiver is developed in \cite{SI-DFT-s-OFDM}. The proposed SensingNet and ComNet can jointly estimate sensing parameters and demodulate data symbols during passive sensing.
Assuming a common distribution of dominant paths for both C\&S channels, \cite{SharedPath} estimates the covariance matrix of the communication channel using echo signals. In this case, the base station (BS) can perform beamforming without requiring any feedback from UEs.
In \cite{IEEE802}, a brute-force optimization algorithm is employed for radar ranging based on IEEE 802.11a/g/p OFDM communication signals, where the mean-normalized channel energy is modeled as a direct sinusoidal function and determined for range estimation.

Nevertheless, considering the generality and synergy of C\&S, the aforementioned schemes still have their limitations:
\begin{itemize}
	\item While capable of achieving communication-assisted sensing or sensing-assisted communication, these approaches have not yet realized synergy through joint processing of the two modalities.
	\item These schemes \cite{QoS, JRCD-SOTA} primarily focus on feedback-free time division duplex (TDD) systems and are not applicable to frequency division duplex (FDD) systems. This limitation arises due to extra large-scale receivers required for FDD systems, leading to unaffordable hardware costs. Furthermore, in the FDD mode as well as the high-frequency TDD mode, perfect uplink/downlink channel reciprocity does not hold, which requires downlink channel state information (CSI) feedback from UEs \cite{CSAB}. However, relevance between communication feedback and sensing has not been considered in these schemes.
\end{itemize}

\section{Proposed Joint Communication and Sensing Channel Semantic Acquisition Scheme}\label{S3}

\begin{figure*}[t]	
	\centering
	{\subfigure[]{
			\includegraphics[width=1.8\columnwidth, keepaspectratio]{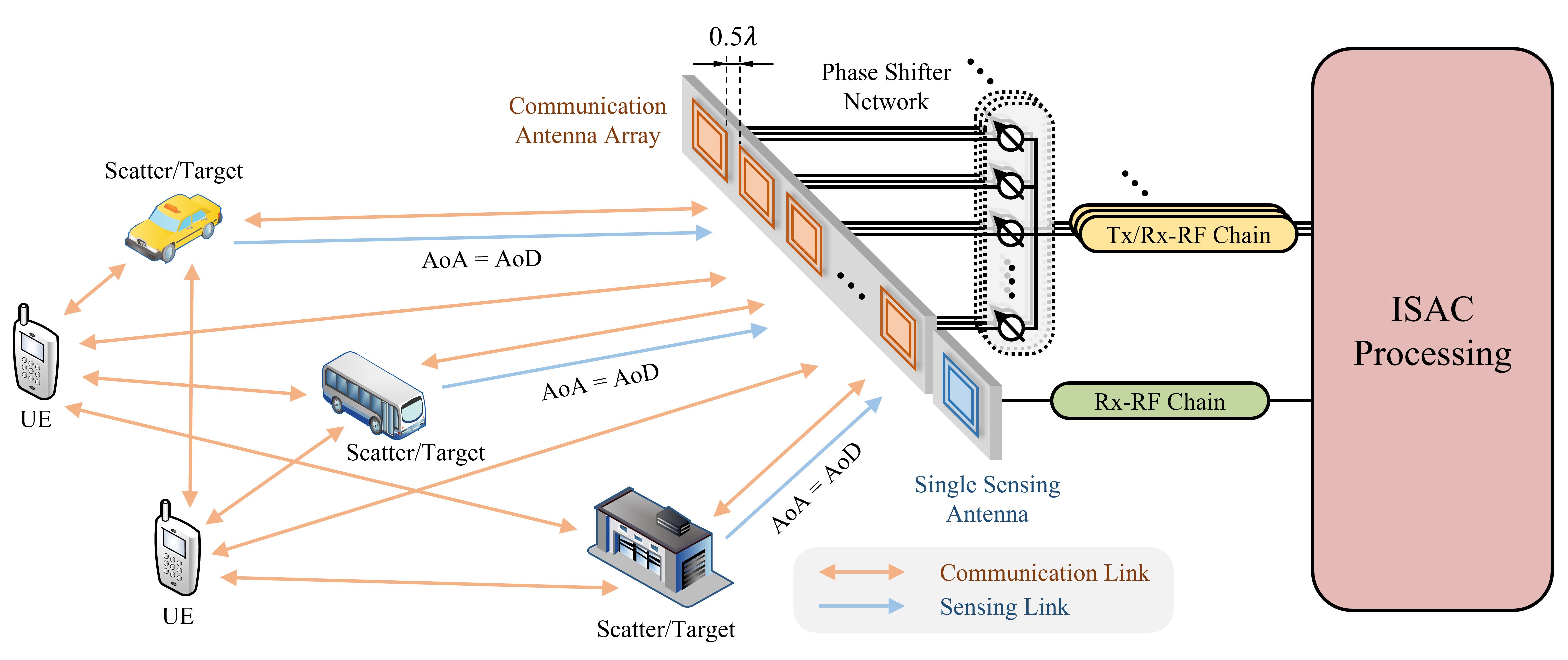}
			\label{hardware_architecture}
		}
		\subfigure[]{
			\includegraphics[width=1.5\columnwidth, keepaspectratio]{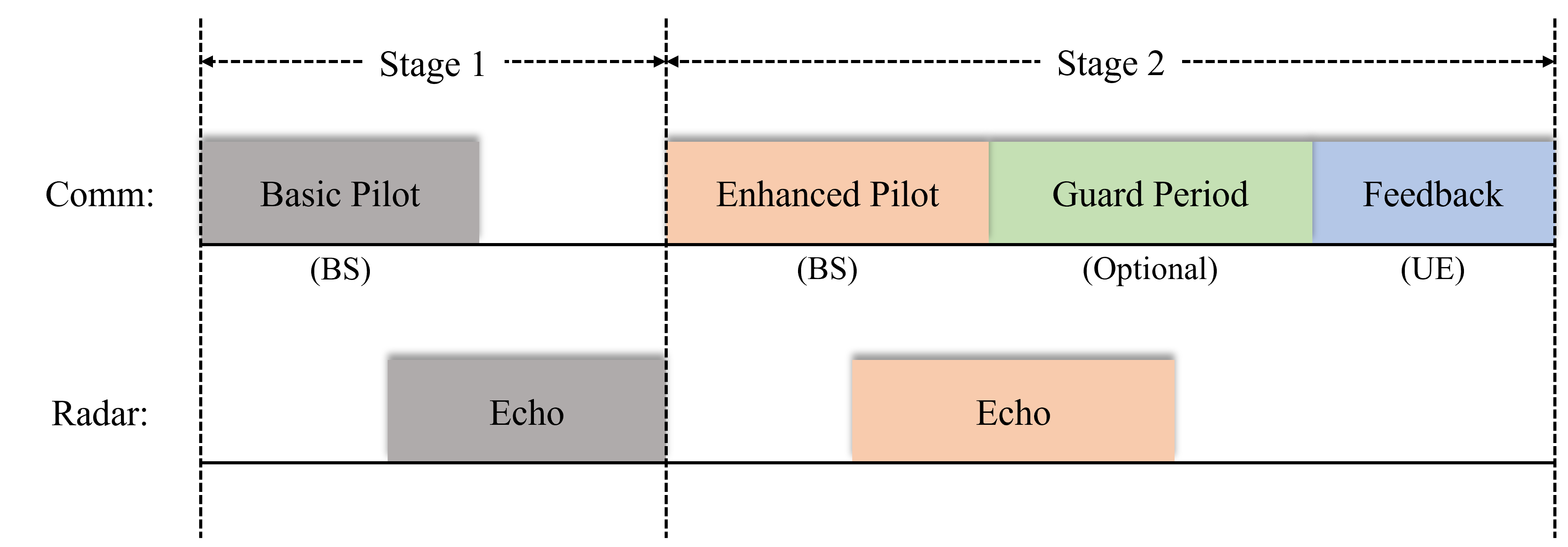}
			\label{frame_structure}
	}}
	\caption{The proposed ISAC system with a single sensing antenna: (a) the hardware architecture of the communication antenna array and the sensing antenna at the ISAC station, and (b) the proposed two-stage frame sutructure.}
	\label{system_framework}
\end{figure*}

To facilitate downlink active sensing, modifications to the hardware of BSs might be required, as the receiver is designed for receiving uplink communication signals, rather than the echoes of downlink communication signals. Such modifications might be unnecessary for TDD systems, since a TDD transceiver employs switches to manage connections between antennas and transmit/receive RF chains. A simple adjustment of the switch connections could potentially suffice for capturing echoes. In contrast, for FDD systems, hardware modifications are necessary as the receiver is incapable of receiving echoes in the downlink frequency bands. Therefore, the implementation of downlink sensing in TDD systems is considered to be more cost-effective \cite{JCR_in_PMN}.
However, due to the limited sensing range in mmWave systems, the overlap between echoes and downlink signals have to be taken into account. Hence in order to integrate sensing capabilities, a TDD system should either achieve full-duplex transmission, which is challenging, or adopts additional sensing antennas.
Moreover, as highlighted in \cite{CS-ISAC}, it has been proved that a receive array tailored for sensing purposes can bring enhanced performance. By employing the HAD-MIMO with low-resolution ADCs, significant reductions in hardware costs and energy consumption can be achieved \cite{JRCD-SOTA, CS-ISAC}.
Furthermore, FDD systems hold the advantage of preventing interference between echoes and uplink communication signals, unlike TDD systems. This advantage allows downlink active sensing in FDD systems to operate without the requirement for guard period between echoes and uplink reception, thereby reducing latency.

Consider an ISAC system shown in Fig. \ref{hardware_architecture}. The BS adopts a uniform linear array with $M$ antennas, where the fully connected HAD-MIMO with $K$ RF chains is employed. Additionally, a single sensing antenna is deployed for downlink active sensing, offering the advantages of minimizing the hardware costs and energy consumption. Furthermore, we consider that the BS simultaneously serves $K$ single-antenna UEs and employs the cyclic prefix (CP)-OFDM with $N_c$ orthogonal subcarriers to combat the frequency-selective fading in broadband transmission.

Compared to purely acquiring communication channel semantic in \cite{CSAB}, the ISAC scheme involving channel semantics of C\&S is more intricate.
We initially divide the process of semantic acquisition for downlink channels into two distinct steps: extraction and reconstruction.
Regarding the channel semantic extraction, we introduce a two-stage frame structure for TDD systems with imperfect RF calibration and FDD systems, as illustrated in Fig. \ref{frame_structure}. The guard period is indispensable in TDD systems to avoid collisions between echos and UEs' feedback, while it can be removed in FDD systems.

\begin{figure*}[!h]
	\centering
	\vspace{-5mm}
	\includegraphics[width=1.8\columnwidth, keepaspectratio]{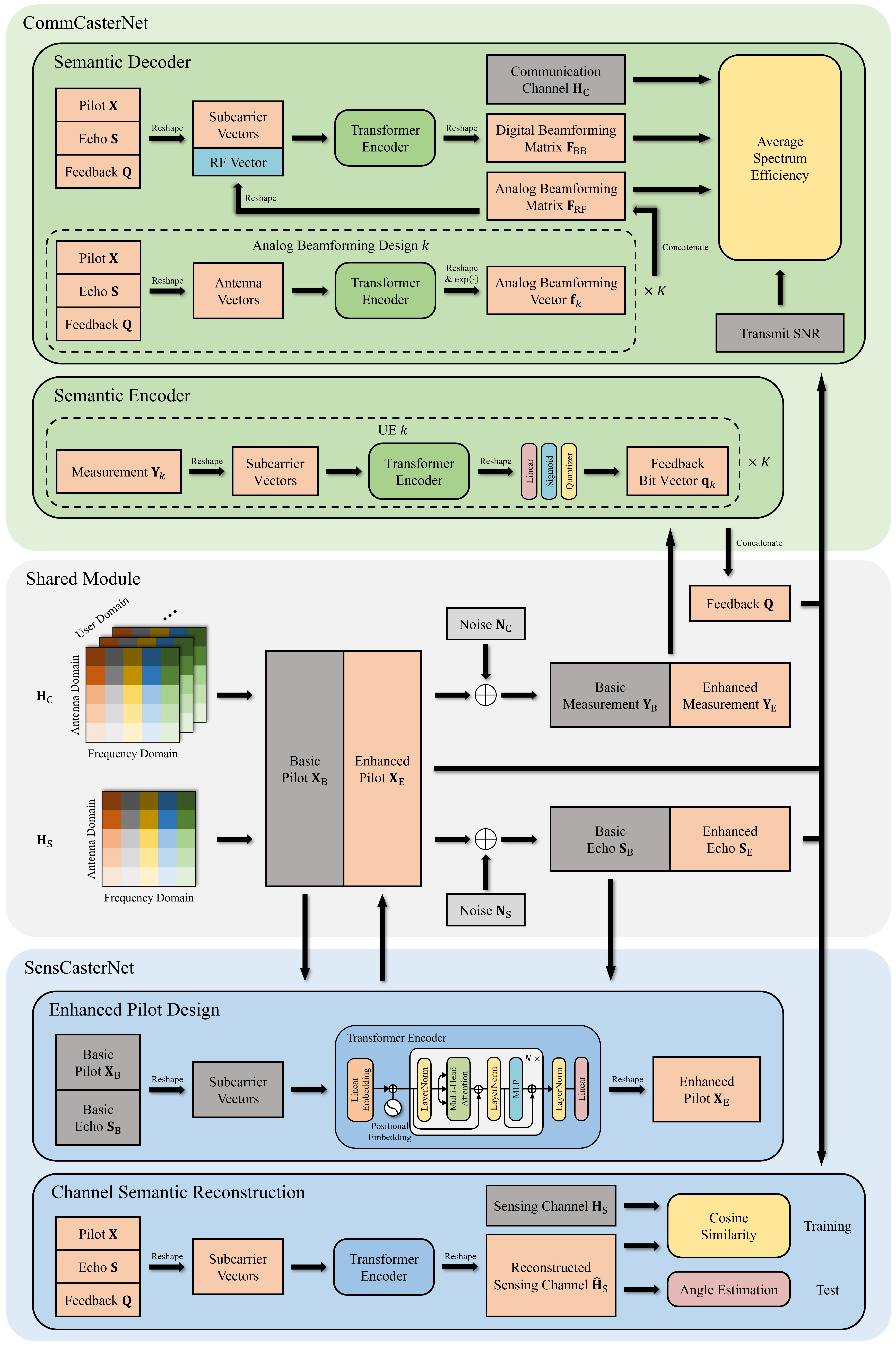}
	\caption{The block diagram of the proposed JCASCasterNet.}
	\label{block_diagram}	
\end{figure*}

In the first stage, the basic pilot is transmitted for rough extraction of the sensing channel semantic, which is then utilized to design the enhanced pilot. Considering that the dominant paths in C\&S channels are partially similar, the enhanced pilot can improve the efficiency of respective channel semantic extraction. Subsequently, UEs extract the communication channel semantic using measurements gathered from both stages and then feed it back to the BS. Finally, the BS performs joint C\&S channel semantic reconstruction by cooperatively processing echoes and UEs' feedback.
To achieve synergy, we propose a DL-based \textbf{j}oint \textbf{c}ommunication \textbf{a}nd \textbf{s}ensing \textbf{c}h\textbf{a}nnel \textbf{s}eman\textbf{t}ic \textbf{e}xtraction and \textbf{r}econstruction network, called JCASCasterNet. The block diagram of the proposed scheme is shown in Fig. \ref{block_diagram}.

JCASCasterNet can be partitioned into three components: CommCasterNet, SensCasterNet, and the shared module. Global optimization is achieved through end-to-end training. Moreover, the allocation of pilot symbols in the two stages, combined with the weighting of loss functions, can accommodate the degree of sharing between communication channel semantic and sensing channel semantic, and the tradeoff between the C\&S performance.

In the following sections, we will provide a detailed exposition of the joint design framework. Subsequently, the effectiveness of the proposed scheme will be demonstrated through an illustrative case study, conducted under significantly limited pilot symbols and feedback signaling overhead.

\subsection{Stage 1: Basic Pilot Based Channel Sounding}

During the first stage with the basic pilot, UEs can acquire downlink communication CSI, and the BS can be regarded as a monostatic radar. The uniqueness lies in the fact that the radar receiver consists of only a single antenna. Leveraging the fact that the angle of departure (AoD) and AoA in monostatic radar are the same, reconstruction of the sensing channel can be formulated as a CS problem. However, the goal of the first stage is to improve the efficiency of channel semantic extraction at the second stage, which can not be achieved by conventional methods. Thus, we conceive a transformer-based approach for enhanced pilot design. Relying on the self-attention mechanism of transformer, the common channel semantic for C\&S is roughly extracted and employed to assist the refinement of the next stage.

Contrasted with the enhanced pilot, the trainable basic pilot, although lacking the adaptability to specific samples, demonstrates generalization ability in diverse scenarios. The increasing overhead of the basic pilot assists the design of the enhanced pilot, promoting the utilization of sensing channel semantic. While in the case that a substantial difference exists between C\&S channels, this would potentially result in limited improvement in communication benefited from sensing. Additionally, the reduction of the enhanced pilot symbols reduces the adaptability of the scheme, which might lead to an overall performance degradation. Yet, conversely, insufficient basic echoes hinders the effective design of the enhanced pilot. Therefore, a proper allocation of pilot symbols at two stages is crucial. It is worth noting that the weights of C\&S loss functions also have an impact on the enhanced pilot design. Due to the partial dissimilarity, such as the line-of-sight (LoS) paths present in the communication channel but are not observable in the sensing channel, power allocation in the angular domain becomes inevitable for the enhanced pilot to ensure the communication performance.

\subsection{Stage 2: Enhanced Pilot Based Distributed Active/Passive Sensing}

Indeed, the sensing antenna at the BS can be viewed as a virtual UE, which can share the received measurements with the BS for the second stage processing. In this manner, during the second stage of channel sounding, the virtual UE and real UEs form a distributed multi-static radar system. Here the real UEs can be regarded as a type of passive sensing receivers. Based on the measurements
of basic and enhanced pilots at two stages, real UEs respectively can extract the communication-exclusive channel semantic and the C\&S common channel semantic, and then feed them back to the BS for ISAC design. In this distributed sensing system, feedback and echoes can be collaboratively processed, resulting in the synergy of C\&S. It is important to emphasize that interference between feedback and echoes needs no consideration in FDD systems, but requires a guard period in TDD systems.

\subsection{Channel Semantic Reconstruction}

In order to achieve synergy, the channel semantic reconstruction at the BS exploits the pilots, echoes, and feedback signals.

For the reconstruction of communication channel semantic, we adopt average spectral efficiency (ASE) per UE (refer to Formula 5 in \cite{DFRC-OWD}) as the loss function. The goal-oriented approach enables UEs to eliminate redundancy and embed the essential semantic required for beamforming into the feedback.
Nevertheless, ASE presents ambiguity in HAD-MIMO, undermining the convexity of the hybrid beamforming problem. To be specific, when simultaneously swapping the columns of the analog beamforming matrix and corresponding rows of the digital beamforming matrix, the ASE performance remains unchanged, which suggests that the channel has multiple optimal hybrid beamforming solutions, thereby disrupting the convergence of neural network.
In light of this, we introduce a knowledge-inspired model design. As demonstrated by beam alignment algorithms, the phase shifters connected to different RF chains can be individually set according to the corresponding UE channel \cite{CSAB}. Based on this idea, feedback from each UE is employed to design a column of the analog beamforming matrix in our method, thus eliminating ambiguity in the analog beamforming problem. Furthermore, considering that analog beamforming primarily leverages information from the angular domain, patches are divided based on the antenna dimension and then fed into the transformer encoder. This design also facilitates the network to generalize to different numbers of UEs and reduces the network parameters. Taking inspiration from the concept of the effective channel matrix, the analog beamforming matrix is then embedded as an additional token input to the digital beamforming design network, effectively resolving ambiguity in the digital beamforming problem.

As for the reconstruction of sensing channel semantic, directly fitting the AoA would bring several issues, e.g., the output ordering has ambiguity, the output dimension varies with the number of targets. In contrast, the channel subspace is relatively simple to reconstruct and contains all the necessary information needed for AoA estimation without redundancy, as proven by algorithms like MUSIC and estimation of signal parameters via rotational invariance techniques (ESPRIT). Given the rotational invariance of the channel subspace, we employ cosine similarity as the loss function. 
Since the downlink AoA estimation does not require the frequency domain information of the sensing channel, different from the commonly used normalized mean square error (NMSE), cosine similarity can concentrate on recovering the channel subspace without being affected by phase variations across different subcarriers.
Moreover, due to the identical channel subspace shared by different subcarriers, the proposed scheme is capable of exploiting this correlation to achieve accurate sensing channel semantic reconstruction even when the number of pilot symbols is less than the target number and the signal-to-noise ratio (SNR) is low.

\subsection{Case Study}

\begin{figure*}[t]	
	\centering
	{\subfigure[]{
			\includegraphics[width=0.65\columnwidth, keepaspectratio]{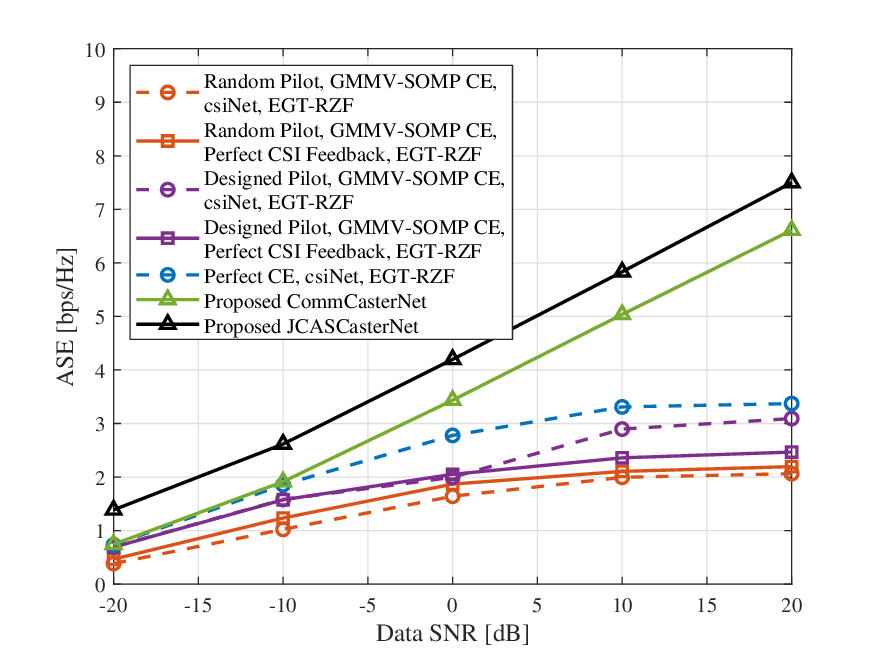}
			\label{Com_ASEvsSNR_PilotSNR=-10_EchoSNR=10}
		}
		\subfigure[]{
			\includegraphics[width=0.645\columnwidth, keepaspectratio]{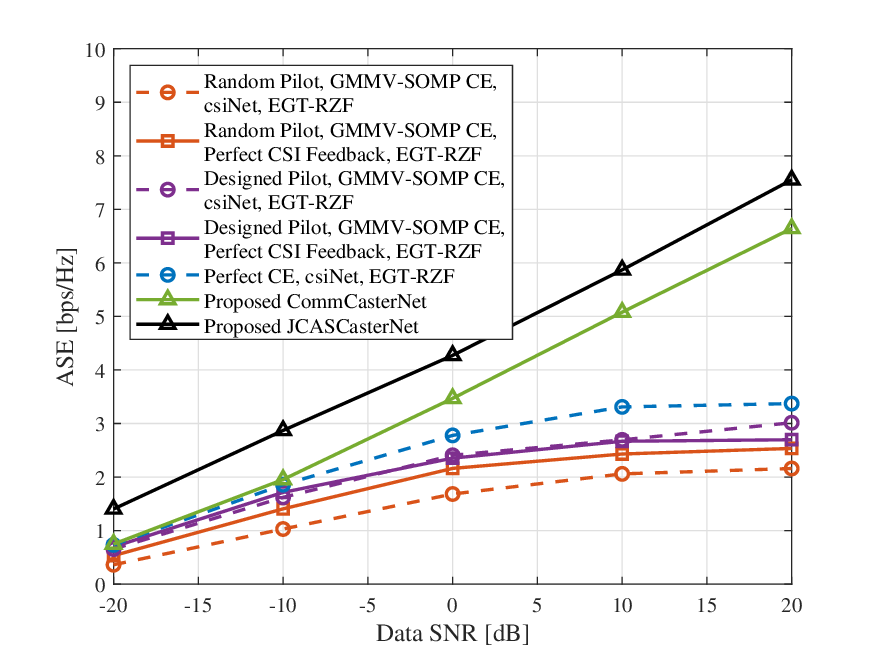}
			\label{Com_ASEvsSNR_PilotSNR=0_EchoSNR=10}
		}
		\subfigure[]{
			\includegraphics[width=0.64\columnwidth, keepaspectratio]{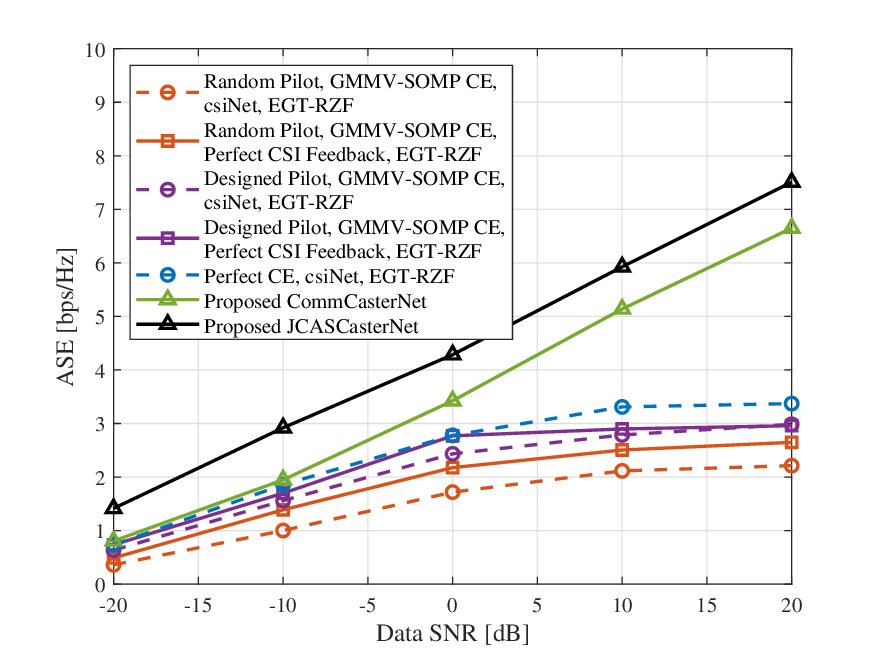}
			\label{Com_ASEvsSNR_PilotSNR=10_EchoSNR=10}
	}}
	\caption{ASE performance comparison versus the data SNR:
		(a) $\textrm{Pilot SNR} = -10 \ \textrm{dB}$;
		(b) $\textrm{Pilot SNR} =   0 \ \textrm{dB}$;
		(c) $\textrm{Pilot SNR} =  10 \ \textrm{dB}$.}
	\label{Com_simulation}
\end{figure*}

\begin{table}[]
	\centering
	\caption{Simulation parameter settings}	
	\label{settings}{%
		\begin{tabular}{|c|c|}
			\hline
			\textbf{Parameter}                      			& \textbf{Value}   			\\ \hline
			Carrier frequency {[}GHz{]}             			& 10               			\\ \hline
			Bandwidth {[}MHz{]}              	     			& 30.72            			\\ \hline
			Number of subcarriers                   			& 32               			\\ \hline
			\multirow{2}*{Communication antenna array size}	& Fully connected HAD-MIMO		\\ 
			~                                              	& with 32 antennas and 2 RF chains	\\ \hline
			Sensing antenna array size					& Single antenna				\\ \hline
			Number of single-antenna UEs					& 2						\\ \hline
			Azimuth angle range $[^\circ]$				& $[-90,90)$	   			\\ \hline
			Coverage radius {[}m{]}						& 100						\\ \hline
			Number of targets							& 6						\\ \hline
			Number of basic pilot symbols					& 2						\\ \hline
			Number of enhanced pilot symbols				& 2						\\ \hline
			Feedback signaling overhead {[}bit{]}			& 16						\\ \hline
		\end{tabular}%
	}
\vspace{-3mm}
\end{table}

To demonstrate the superiority of the proposed scheme, we study a representative case in the massive MIMO-based ISAC systems. Generalized multiple-measurement-vectors (GMMV)-simultaneous orthogonal matching pursuit (SOMP) \cite{GMMV-SOMP} is employed as a benchmark for CS algorithms, as optimization-based iterative algorithms converge slowly in this case. Following the reconstruction of the sensing channel, angle estimation is accomplished using the MUSIC algorithm. Besides, the DL-based csiNet \cite{csiNet} is adopted as the baseline solution for downlink communication CSI feedback and reconstrcution. With the reconstructed communication channels, the benchmark schemes employ equal gain transmission (EGT) and regularized zero forcing (RZF) to design the downlink hybrid beamforming \cite{CSAB}. Ablation experiments are conducted using CommCasterNet and SensCasterNet separately without mutual information to demonstrate the synergy of C\&S in JCASCasterNet. To validate the effectiveness of the frame structure and the enhanced pilot, the random pilot is utilized as a benchmark for comparison.
The simulation parameter settings are summarized in TABLE \ref{settings}.
To highlight the correlation between C\&S channels, we assume that the scatterers of the communication channel are all regarded as targets in the sensing channel, and each UE has a LoS path. The SNR used in the case study is all referred to as transmit SNR.

Fig. \ref{Com_simulation} shows the ASE performace achieved by different schemes versus the data SNR at different pilot SNRs. For the JCASCasterNet, the echo SNR is set to 10 dB. It can be observed that the designed pilot (includes basic pilot and enhanced pilot) performs better than the random pilot, demonstrating the effectiveness of the proposed frame structure and the enhanced pilot for communication. Moreover, the proposed methods outperform ``Perfect CE, csiNet, EGT-RZF'', confirming the superiority of channel semantic acquisition techniques. On the other hand, schemes with ``GMMV-SOMP CE, Perfect CSI Feedback, EGT-RZF'' is inferior to ``Perfect CE, csiNet, EGT-RZF'', validating that the bottleneck of the baseline methods largely stems from the limited number of pilot symbols. While CommCasterNet manages to partially mitigate this issue via channel semantic acquisition, JCASCasterNet consistently shows superior performance with around 1 bps/Hz. This result underscores the benefit of integrating sensing capabilities to enhance communication performance. Besides, all the schemes exhibit relatively consistent ASE performance at different pilot SNRs, indicating that pilot and feedback signaling overhead plays a dominant role in the performance of hybrid beamforming design. The proposed channel semantic acquisition techniques remain optimal even at a low pilot SNR, highlighting robustness of the channel semantic to noise.

\begin{figure*}[t]	
	\centering
	{\subfigure[]{
			\includegraphics[width=0.9\columnwidth, keepaspectratio]{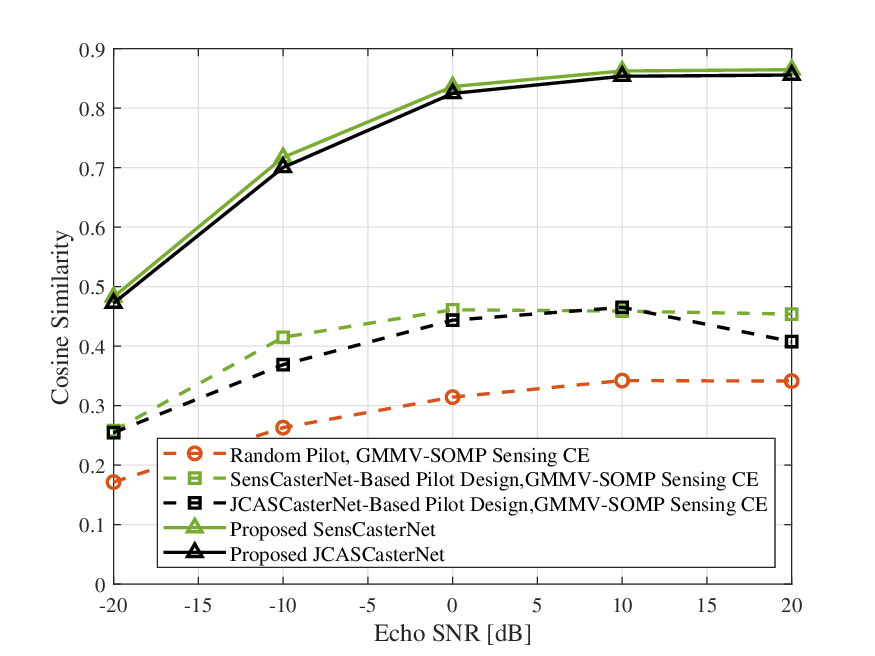}
			\label{Rad_CosvsSNR_PilotSNR=10_SNR=10}
		}
		\subfigure[]{
			\includegraphics[width=0.9\columnwidth, keepaspectratio]{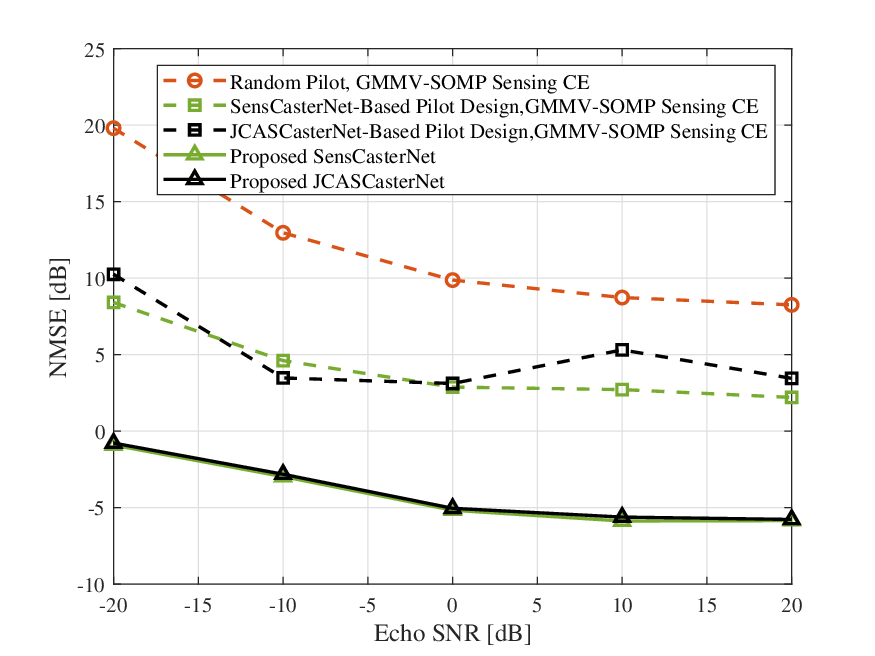}
			\label{Rad_NMSEvsSNR_PilotSNR=10_SNR=10}
	}}
	\caption{Sensing channel semantic reconstruction performance with different loss functions:
		(a) Cosine Similarity; (b) NMSE.}
	\label{Rad_simulation1}
\end{figure*}

\begin{figure*}[t]	
	\centering
	{\subfigure[]{
			\includegraphics[width=0.9\columnwidth, keepaspectratio]{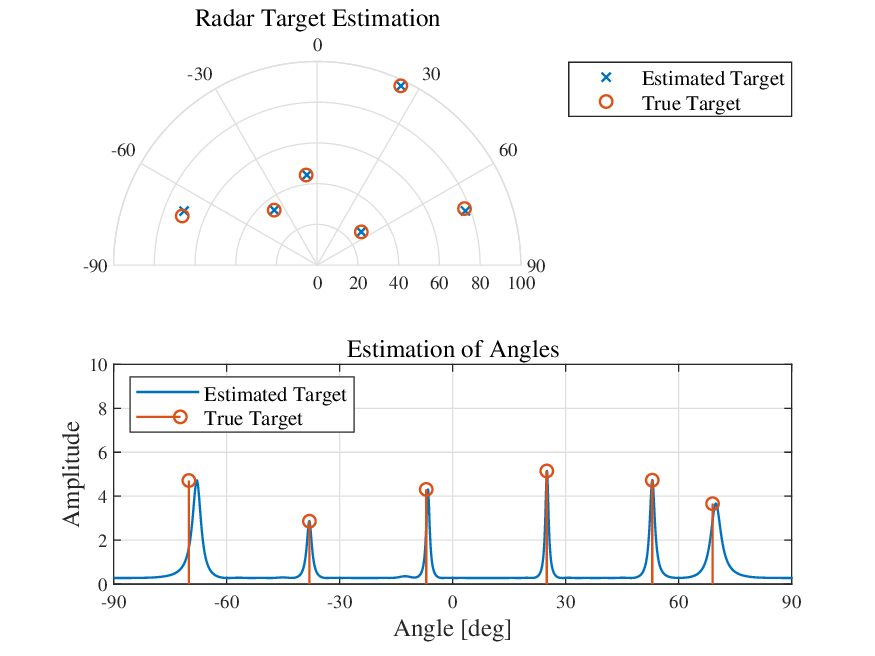}
			\label{MUSIC_Cos_PilotSNR=10_EchoSNR=0_SNR=10}
		}
		\subfigure[]{
			\includegraphics[width=0.9\columnwidth, keepaspectratio]{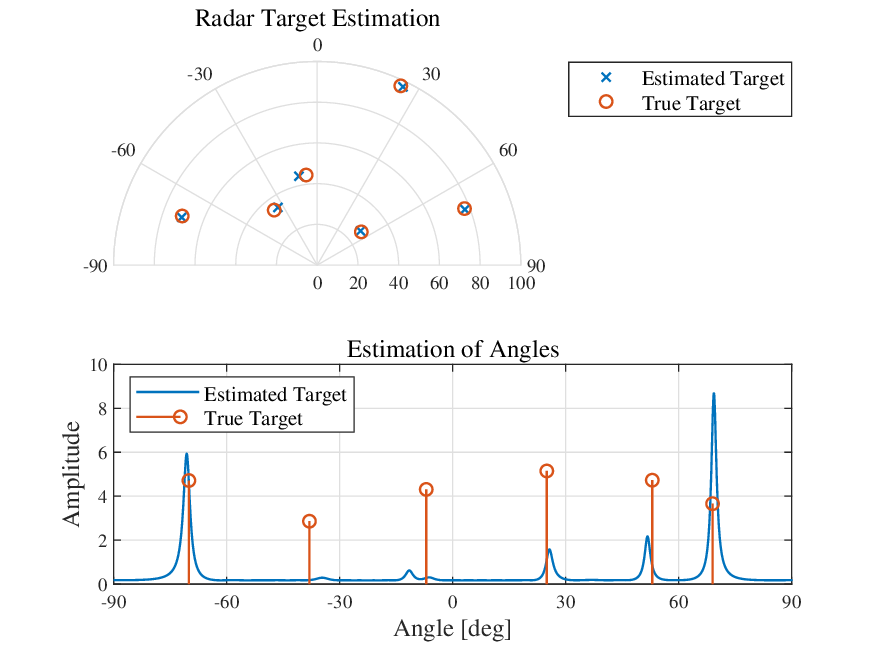}
			\label{MUSIC_NMSE_PilotSNR=10_EchoSNR=0_SNR=10}
	}}
	\caption{Angle estimation performance comparison of different loss functions:
		(a) Cosine Similarity; (b) NMSE.}
	\label{Rad_simulation2}
\end{figure*}

Fig. \ref{Rad_CosvsSNR_PilotSNR=10_SNR=10} and Fig. \ref{Rad_NMSEvsSNR_PilotSNR=10_SNR=10} respectively depict the cosine similarity and NMSE as a function of echo SNR. For the JCASCasterNet, the pilot SNR is set to 10 dB. Constrained by the limited number of pilot symbols, the benchmarks suffer a significant performance loss regardless of the echo SNR. However, the approaches with the enhanced pilot still outperforms those utilizing the random pilot, reconfirming the effectiveness of the proposed two-stage frame structure for sensing. Furthermore, the SensCasterNet-based pilot design achieves better performace than the JCASCasterNet-based pilot design. This confirms the earlier analysis that power allocation in the angular domain becomes inevitable for the enhanced pilot to ensure the communication performance because of the partial dissimilarity between communication amd sensing channels. The enhanced pilot designed in JCASCasterNet cannot dedicate to sensing like SensCasterNet, which consequently leads to the degraded performance in sensing channel semantic reconstruction. Nevertheless, with the assistance of UEs' feedback, the gap between the sensing performance of JCASCasterNet and SensCasterNet remains minimal.
Fig. \ref{Rad_simulation2} provides additional insight into the performance of channel subspace reconstruction with different loss functions at an echo SNR of 0 dB. By transmitting different pilots on different subcarriers, the proposed approaches capitalize on the identical channel subspace across different subcarriers so that the angular information can be reconstructed with fewer pilot symbols than the target number. In the scheme with NMSE as the loss function, the orthogonality between the channel subspace and noise subspace is disrupted, resulting in the angular spectrum at certain positions being nearly overwhelmed by noise. In contrast, cosine similarity enables a more precise reconstruction of the sensing channel semantic.

\section{Open Issues and Future Directions}\label{S4}

DL-based channel semantic acquisition offers opportunities for ISAC networks to realize synergy, yet the progress still confronts a range of challenges. In this section, we succinctly outline some open issues and promising directions for ISAC networks.

\subsection{Channel Modeling of ISAC Networks}

As mentioned above, leveraging the correlation between C\&S channels can reduce the pilot and feedback signaling overhead, enable seamless synergy between C\&S, and ultimately improve the overall performance of ISAC networks. However, it remains unclear how to accurately model this correlation in practical environment. In the current stage, the common angular information of objects serving as both communication scatterers and sensing targets has been widely adopted, yet discussions about other crucial factors such as cluster structure and energy distribution are relatively scarce. Hence, the lack of precise channel models of ISAC networks has limited the exploration of synergy schemes.

\subsection{Synergy Performance Analysis under Explainable AI}

To gain a deeper comprehension of ISAC systems, synergy performance analysis is essential. Furthermore, the current training approach of weighting the C\&S loss functions is still rather rudimentary. Reliable synergy performance analysis can also provide more accurate objectives for DL-based ISAC schemes, guide the design of network architectures, and improve the explainability of AI. Although the information theory of ISAC has yielded some preliminary results, many practical constraints have not been taken into consideration. With the analysis of synergy performance, it is believed that we can attain deeper insight into the criteria of AI empowered ISAC networks and accomplish a higher level of synergy.

\subsection{Lightweight Deployment and Multi-BS Sensing}

AI technology, while boosting the performance of ISAC networks, imposes increased demands on computational resources. To reduce hardware costs and inference overhead, further research on lightweight deployment for ISAC networks, such as quantization, pruning, knowledge distillation, etc., is imperative. Additionally, this paper only considers ISAC design with a single BS, but multi-BS sensing is also important in practical scenarios. For instance, when multiple targets are detected within the sensing coverage of several BSs simultaneously, target association becomes necessary to determine which parameters sensed by different BSs correspond to the same target. In order to implement ISAC technology, lightweight deployment and multi-BS sensing are indispensable.

\subsection{Security Issues of ISAC Networks}

With the emergence of 6G and its intelligent applications such as remote-Health, V2X, and smart home, a substantial amount of personal data is being delivered, giving rise to concerns about the security of ISAC networks. Sensing requires the transmitted waveforms to thoroughly interact with the wireless environment, rather than just serve the UEs in communication. The interaction not only embeds target parameter information into echoes but also poses risks for eavesdropping and privacy leakage. Currently, research on the security of ISAC networks is still in its early stages. Striking a balance between security and efficiency stands as a pivotal topic for future explorations.

\section{Conclusions}\label{S5}

In this paper, we introduce an AI-based channel semantic acquisition framework for ISAC networks, achieving synergy between C\&S. To begin with, we conduct a comprehensive overview of the state-of-the-art solutions encompassing waveform design and signal processing, while also discussing their inherent limitations. Considering the correlation between C\&S channels, we propose a two-stage frame structure along with the channel semantic learning based JCASCasterNet, which yield gains in both extraction and reconstruction of channel semantic. The case study demonstrates that the proposed scheme achieves improved overall performance even with extremely low pilot and feedback signaling overhead. Finally, we highlight the challenges that ISAC networks encounter in its development and outline potential research directions in the upcoming 6G era.

\end{document}